\documentclass{elsart}
\usepackage{graphicx}

\include{elsart.cls}

\begin{document}

\begin{frontmatter}
\title{Characterization and Modeling of Non-Uniform Charge Collection
in CVD Diamond Pixel Detectors}

\author[Milano,Bonn]{T. Lari\thanksref{TL}}, 
\author[CERN]{A. Oh},
\author[Bonn]{N. Wermes},
\author[Ohio]{H. Kagan},
\author[CERN,Bonn]{M. Keil},
\author[Toronto]{W. Trischuk}

\address[Milano]{INFN and Dipartimento di Fisica, Universit\`a di Milano, 
                 Via Celoria 16, I-20133, Milano (Italy)}
\address[Bonn]{Physikalisches Institut der Universit\"{a}t Bonn,
                 Nussallee 12, 53115 Bonn, Germany}
\address[CERN]{European Laboratory for Particle Physics (CERN),  
                 1211 Geneva 23, Switzerland}
\address[Ohio]{High Energy Physics Group, Department of Physics, 
               Ohio State University, 174 West 18th Avenue, Columbus, OH, USA}
\address[Toronto]{Department of Physics,University of Toronto, 
                  60 St. George Street, Toronto, ON, Canada}

\thanks[TL]{Corresponding author. Tel: +39-0250317391; fax: +39-0250317624. 
{\em E-mail address} Tommaso.Lari@mi.infn.it}

\begin{abstract}

A pixel detector with a CVD diamond sensor has been studied in a 180 GeV/c
pion beam.
The charge collection properties of the diamond sensor were studied  as a 
function of the track position, which was measured with a silicon microstrip 
telescope.
Non-uniformities were observed on a length scale comparable to the 
diamond crystallites size. In some regions of the sensor, the charge drift
appears to have a component parallel to the sensor surface (i.e., normal 
to the applied electric field) resulting in systematic residuals between 
the track position and the hits position as large as 40~$\mu$m.  
A numerical simulation of the charge drift in polycrystalline diamond 
was developed to compute the signal induced on the electrodes by the 
electrons and holes released by the passing particles. The simulation 
takes into account the crystallite structure, non-uniform trapping across 
the sensor, diffusion and polarization effects. It  
is in qualitative agreement with the data.
Additional lateral electric field components result from  the
non-uniform trapping of charges in the bulk. These provide a good 
explanation for the large residuals observed.
\end{abstract}


\end{frontmatter}

\section{Introduction}

In the last decade the quality of
chemical vapor deposited (CVD) diamond for particle detectors has greatly 
improved, and 
the use of this material has become a potentially attractive option 
for vertex detectors at
high luminosity colliders (such as the LHC after the proposed luminosity 
upgrade~\cite{Gia02}) promising to provide the
radiation resistance needed for the challenging particle fluxes expected.

Diamond detectors are believed to be more radiation-hard than 
silicon~\cite{Mei99,Ada00,Lin01}.
Even after radiation exposures in excess of  
$10^{15} \;\; \mbox{hadrons/cm}^2$ they can be operated at room 
temperature without significant leakage current. A loss in the charge 
signal of about 40\% was observed with pion/proton radiation of 
$5 \times 10^{15} \;\; \mbox{p/cm}^{2}$~\cite{Mei99,Ada00} 
. With neutrons a loss of 30\% was observed at a fluence of 
$2 \times 10^{15} \;\; \mbox{n/cm}^{2}$~\cite{Ada00}.

The charge collection distance is the figure of merit for 
diamond particle detectors. It is defined as $d Q/Q_0$ where $d$ is the 
diamond film thickness, and $Q/Q_0$ is the collection efficiency of the 
ionization charge. The collection distance has been increased from 
a few micrometers in the beginning of diamond R\&D 
to about 250~$\mu$m, corresponding to a mean signal of 9000 electrons, 
today~\cite{rd42.02}.

The charge collection distance depends on the presence of charge carrier 
traps~\cite{Zha94}. 
Once a deep level trap has captured a charge carrier the trap can be
permanently passivated. As a consequence of this process, CVD diamond shows 
a significant increase in the charge collection distance after being exposed 
to fluences of the order of $10^{6}$ minimum ionizing particles per 
$\mbox{mm}^2$, a process called pumping or priming. CVD diamond remains 
for a long period (months) in the primed state if kept in the dark and at 
room temperature. 

CVD diamond usually has a polycrystalline structure with 
an average crystallite 
size on the growth side of the order of 1/7 of the thickness~\cite{Ada00}.
On the substrate side the crystal size is only a few micrometers. The 
charge collection distance can be different in different 
crystallites~\cite{Ada00,Oh99}, leading 
to a non-uniform response and a broadening of the 
distribution of the charge signal from the whole substrate.
 
In this paper we report test-beam measurements and 
simulations of diamond pixel detectors equipped with a prototype 
electronics chip of the ATLAS Pixel silicon detector~\cite{Ack99,Bla00,Fis01}.
In Section~\ref{sec2} the analysis of test-beam data taken with 
two diamond pixel detectors is presented. The spatial resolution and 
detection efficiency of the detectors are reported,  
and the observation of a non uniform response of the detectors 
across the sensor area is discussed.  
In Section~\ref{sec3} a numerical simulation of the charge collection 
processes inside the diamond sensor and of the detector response is 
presented. It is shown that the non uniform response can be reproduced 
as a consequence of the polarization fields created by the 
charge trapped in the crystallite structure.

\section{Test beam Set-up and tested devices}

\label{sec2}

\subsection{The test beam set-up}

Test beam experiments were performed at the CERN SPS accelerator 
during the years 
2000 and 2001 with a pion beam of 180 GeV/{\em c} momentum. 

A beam telescope consisting of 4 silicon microstrip modules  
was used to measure the transverse position of the incident beam 
particles. In the setup used in the year 2000~\cite{Lar01} each module 
consisted of a pair of microstrip detectors, each providing one 
coordinate. In the setup used in 2001 each module 
consisted of a double-sided microstrip detector. The new telescope used 
a readout architecture including zero suppression~\cite{Tre02}, 
which offered higher trigger rate capability.  For both set-ups,  
the position resolution of tracks projected onto the tested 
devices was about 6 $\mu$m.

\subsection{The diamond sensors}

Two diamond sensors, identified by UTS-5 and CD91,
were tested in the beam~\cite{Kei03}. The sensors were grown to a thickness 
of about $800 \; \mu$m. The grain size on the growth side is of the 
order of 100~$\mu$m. The sensors were then lapped on the substrate side 
by about $300 \; \mu$m and on the growth side by about 50~$\mu$m 
so that the final thickness is 432~$\mu$m for UTS-5 and 470~$\mu$m 
for CD91. The lateral dimensions of the sensors were $8~\mbox{mm} 
\times 8 \mbox{mm}$.

The electrode on the growth side was segmented in pixels of 
$50 \;\; \mu\mbox{m} 
\times 400 \;\; \mu\mbox{m}$ dimensions. Each pixel was electrically connected 
via electroplated PbSn solder bumps 
to a matching readout electronics cell in the front-end 
chip.

In the following, $x$ is the coordinate along the short (50 $\mu$m)
dimension of the pixels, $y$ is the coordinate along the long 
(400~$\mu$m) dimension of the pixels, and 
$z$ is the coordinate perpendicular to the pixel plane.

Before operation at the test beam, data were taken with a $^{241}$Am
and a $^{90}$Sr source, with activities of 74~MBq and 62~kBq
respectively. The sources were kept for 12~h at 
about 5~mm from the diamond sensor. 
%
%
The $\beta$ source delivered a fluence of the order 
of $10^{7} \; \mbox{mm}^{-2}$ 
(0.3~Gy dose) to each sensor, which is enough to bring good quality diamond 
in the primed state~\cite{Oh99}. This conclusions is supported by the test 
beam data, discussed below, since the average signal and the detection 
efficiency did not
show any time dependence during the data taking.

\subsection{Front-end electronics}

The electronics chips were produced during the development
phase of the ATLAS Pixel front-end electronics 
program~\cite{Ack99,Bla00,Fis01}. Their design was similar to that of 
the final front-end electronics for ATLAS Pixel~\cite{Bla02,Bla04}.
In each front-end chip, 2880 channels 
are arranged into 18 columns by 160 rows.  
The charge-sensitive 
preamplifiers feature a DC feedback scheme with a tunable current 
providing control over the shaping-time for a given input charge. 
A discrimination stage sits behind the preamplifier in each channel which is 
sensitive to the leading edges and the trailing edges of pulses.
Each channel is equipped with its own 3-bit DAC for channel-to-channel 
threshold adjustments, thus a means of overall dispersion reduction is 
provided. The chips have a 7-bit charge measurement capability using 
the time-over-threshold (ToT) of the signal. The ToT is calibrated 
by injecting a known charge into every channel.
A 2880-bit pixel register plus one corresponding latch per channel 
enable individual pixels to be masked-off independently for 
calibration-strobing and readout.

The chip is operated at the 40~MHz LHC bunch crossing rate and the 
times are measured in multiples of the 25~ns clock period.
Only pixel signals whose leading edge belongs to the period 
specified by an external trigger are read out. At the test-beam 
the trigger was provided by two scintillator detectors and 16 consecutive 
25~ns time windows were accepted. 

During the operation in the test beam the thresholds of the individual 
channels were adjusted achieving a threshold dispersion of less than 100~e.
A typical threshold setting was 1000~e while the 
average noise per pixel was about 200~e.

\section{Analysis of test beam data}
\label{analysis}

Events were filtered with the requirement of one and only one track 
reconstructed by the silicon microstrip telescope in each event. 
Only events with a track reconstruction $\chi^2$ probability greater 
than 0.02 were kept. Tracks were required to extrapolate into a fiducial 
area. For CD91 this was the surface covered by the pixel array, excluding 
the region within 40~$\mu$m of the border. 
For UTS-5 it was the smaller area with a good bump-bonding yield~\cite{Kei03}.
 
Pixel clusters were built clustering 
together all adjacent pixels, independently of track extrapolations.
The cluster charge is defined as the sum of the charges measured by 
the pixels of the cluster. This can be less than the total collected charge, 
since it does not include the pixel signals below the electronics threshold.
The cluster position is computed as the arithmetic mean of the  
coordinates of the pixels in the cluster.

The data discussed here have been taken at normal incidence. The sensors 
were operated with a bias voltage of 450~V (UTS-5) or 470~V (CD91).

At normal incidence, most clusters are composed of one or two pixels. 
The average cluster size was measured to be $1.260 \pm 0.001$ pixels 
for CD91 and $1.484 \pm 0.010$ for UTS-5. 

Charge calibrations were available only for UTS-5, for which the cluster
charge (the sum of the charges measured by the pixels of the cluster) 
was $(3000 \pm 300)$~electrons. 
The detection efficiency was $(67.69 \pm 0.10)$\% for CD91 
and $(76.8 \pm 0.5)$\% for UTS-5.  

\begin{figure}[!p]
\begin{center}
\includegraphics[width=10cm,height=8cm]{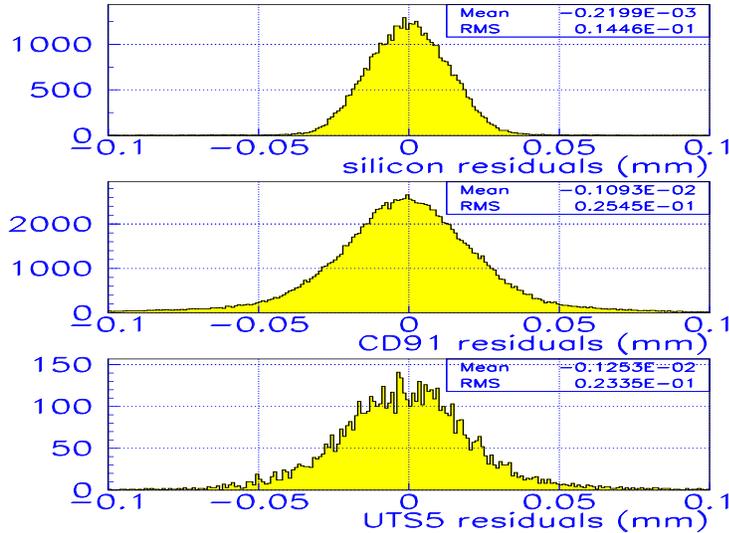}
\end{center}
\caption{\label{residuals}
Residuals between the positions measured by the telescope and by the 
pixel detector for a silicon detector (top), the CVD-diamond sensor 
CD91 (center) and UTS-5 (bottom).}
\end{figure}

Fig.~\ref{residuals} shows the distribution of the $x$ residuals between 
the track position as determined by the telescope and the pixel cluster 
position. For comparison, the same distribution is reported for an ATLAS 
Pixel silicon sensor, with a similar pixel geometry, bump-bonded to a similar 
electronics chip and tested with the same test-beam setup. 
The r.m.s. of spatial residual 
distributions\footnote{The r.m.s. of each distribution is computed 
between -0.1~mm and 0.1~mm} yields  
resolutions of $(14.46 \pm 0.05) \;\mu$m for the silicon detector, 
$(23.35 \pm 0.21)\;\mu$m for UTS-5 
and $(25.45 \pm 0.11) \;\mu$m for CD91. 
The usage of charge interpolation algorithms 
for the position of multi-pixel clusters improves these values only 
by a few tenths of a micrometer. The values of mean cluster size, detection 
efficiency and spatial resolution are summarized in table~\ref{DataTab}.

\begin{figure}[!ht]
\begin{center}
\includegraphics[width=10cm,height=6cm]{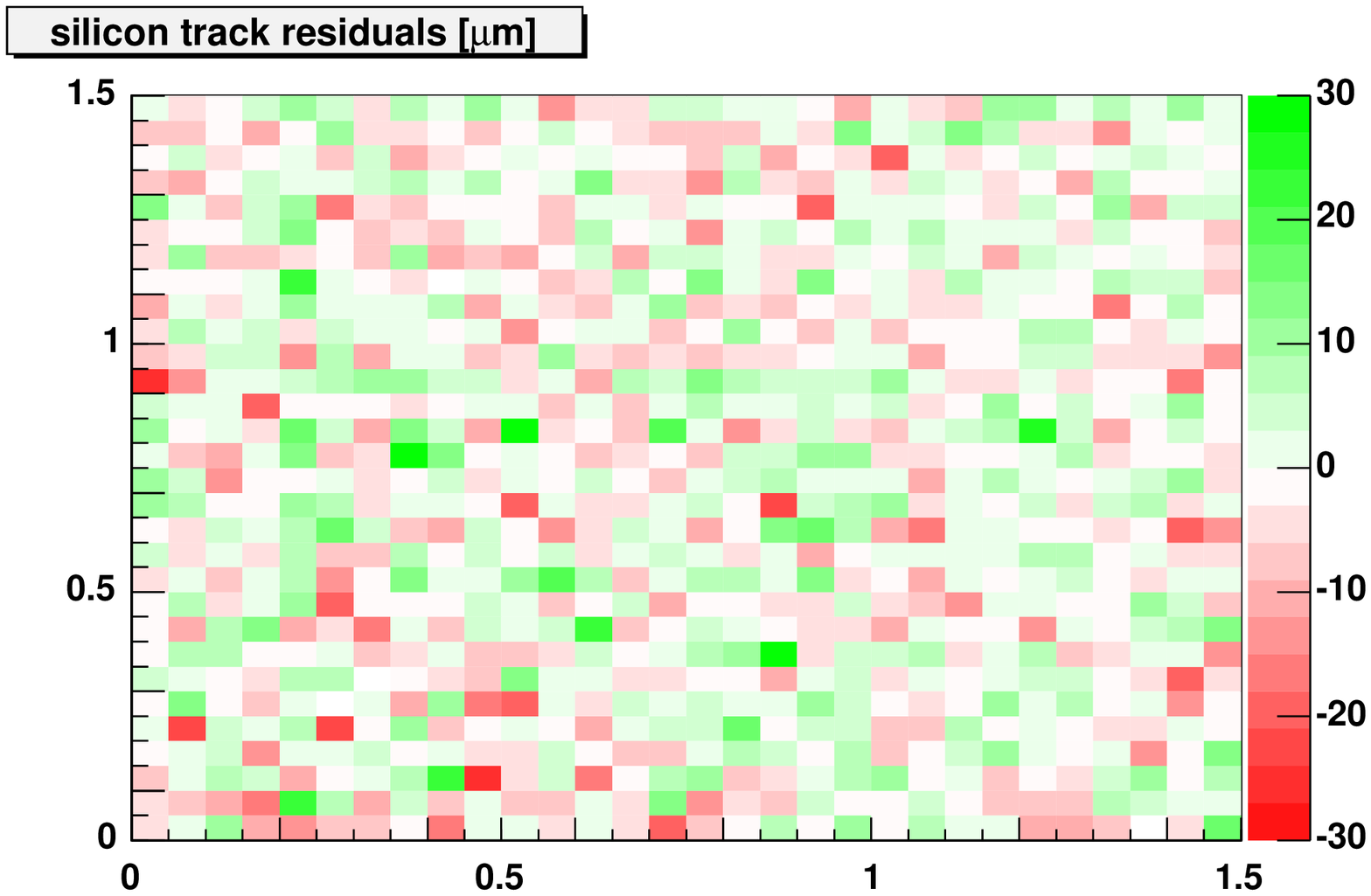}
\includegraphics[width=10cm,height=6cm]{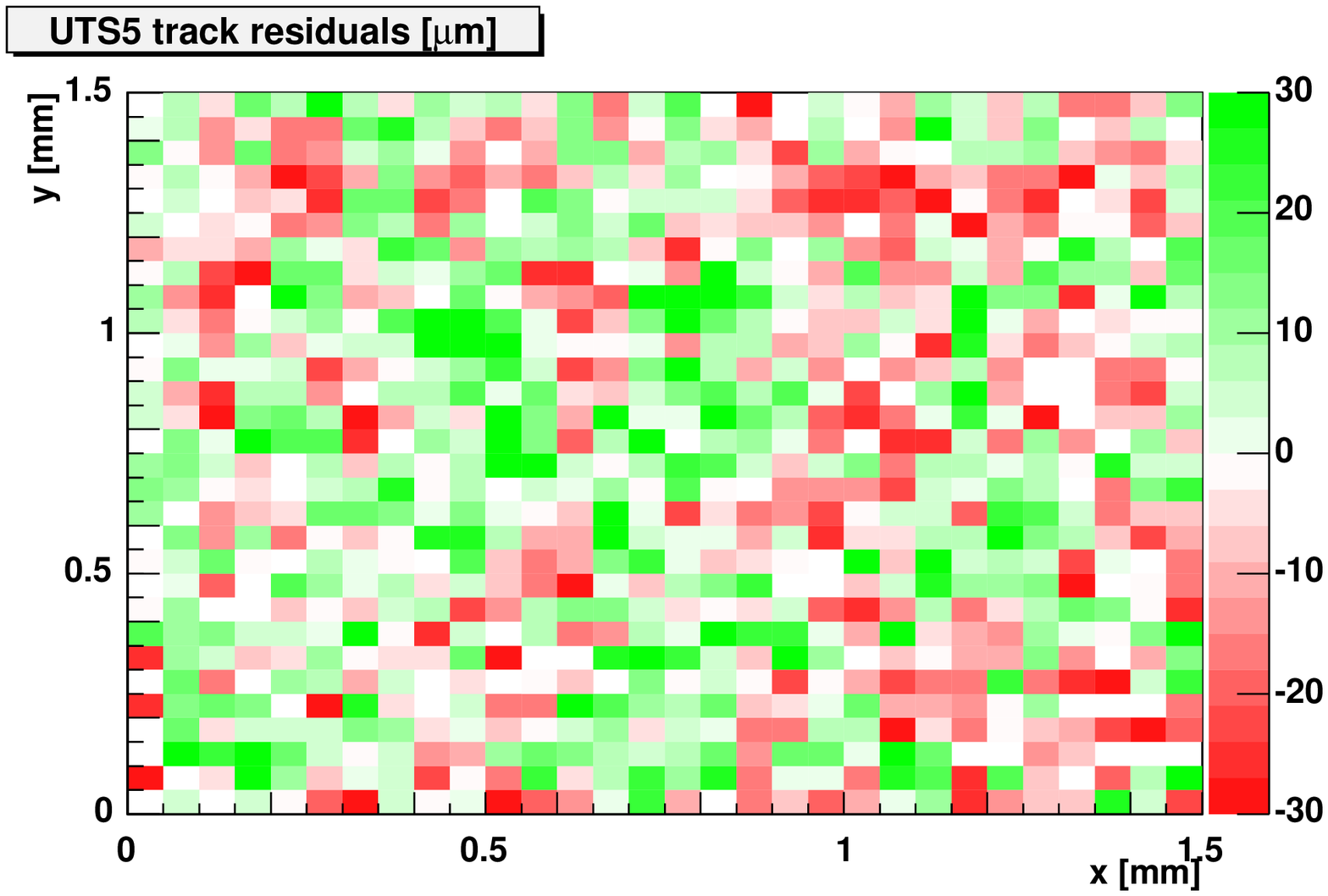}
\includegraphics[width=10cm,height=6cm]{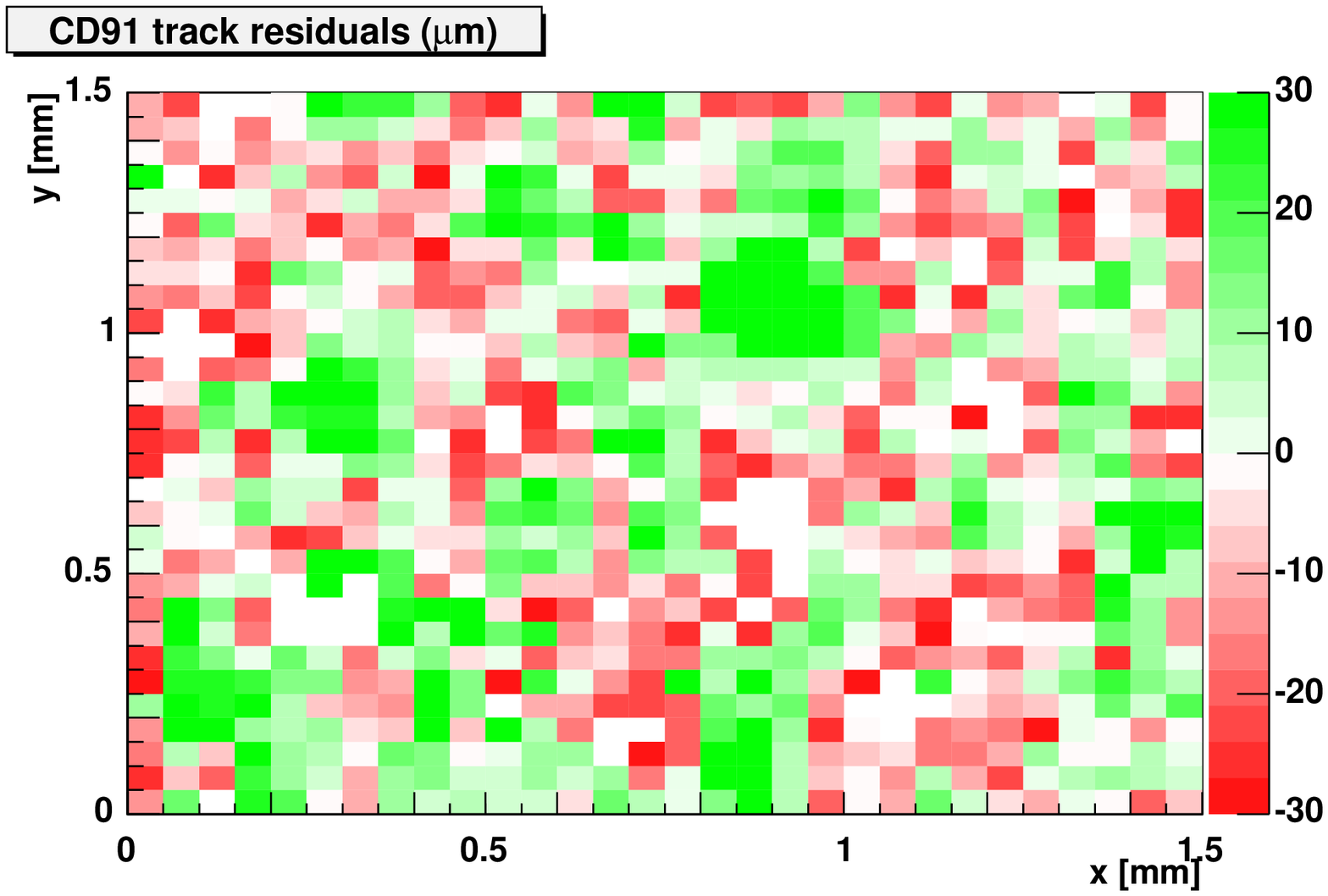}
\end{center}
\caption{\label{ResidualsMap} 
Mean spatial residual between the position of the center of the 
pixel clusters and the track position determined by the tracking 
telescope, as a function of position inside the diamond sensor. 
The upper plot is for a silicon detector, the 
middle one for the diamond sensor UTS-5 and the lower plot is for 
the diamond sensor CD91.} 
\end{figure}

The reasons for the poor diamond resolution were investigated by looking 
at the mean $x$ residuals\footnote{The silicon microstrip telescope provides 
comparable resolutions on the $x$ and $y$ coordinates of the track 
extrapolation. The pixel cluster position, in contrast, is determined 
with far better precision in the $x$ direction, because of the 
8:1 aspect ratio of ATLAS Pixels. For this reason, the $y$ 
residuals have not been taken into consideration.}
as a function of the track position on the sensor. 
The sensor was divided in bins 
of $50 \; \mu\mbox{m} \times 50 \; \mu\mbox{m}$ size, and for each bin $i$ 
the mean spatial $x$ residual $r_i$ has been computed. The same 
analysis was also performed on a ATLAS Pixel silicon detector with similar 
sensor geometry, electronics and test beam setup. 
The results are reported in Fig.~\ref{ResidualsMap}
\footnote{Only a small part of the 
sensor area is covered, to improve the visibility of the residual structures.}.

The silicon sensor (upper plot) shows a more uniform distribution. 
Bins with large 
residuals are isolated and scattered across the sensor as one expects from 
statistical fluctuations. The diamond sensors (middle and lower plot) present  
regions which have systematically positive or negative residuals 
as large as 40~$\mu$m. No correlation is observed between these 
regions and the segmentation in pixels. 
The scale of the residual clustering is of the order of 100~$\mu$m 
which is also the typical size of the diamond crystallites. 
This suggests that in some crystallites the drift of the charge carriers 
has a component parallel to the sensor surface, so that the average position 
of the pixel clusters is shifted away from the track position.   

\begin{figure}[!ht]
\begin{center}
\includegraphics[width=10cm,height=6cm]{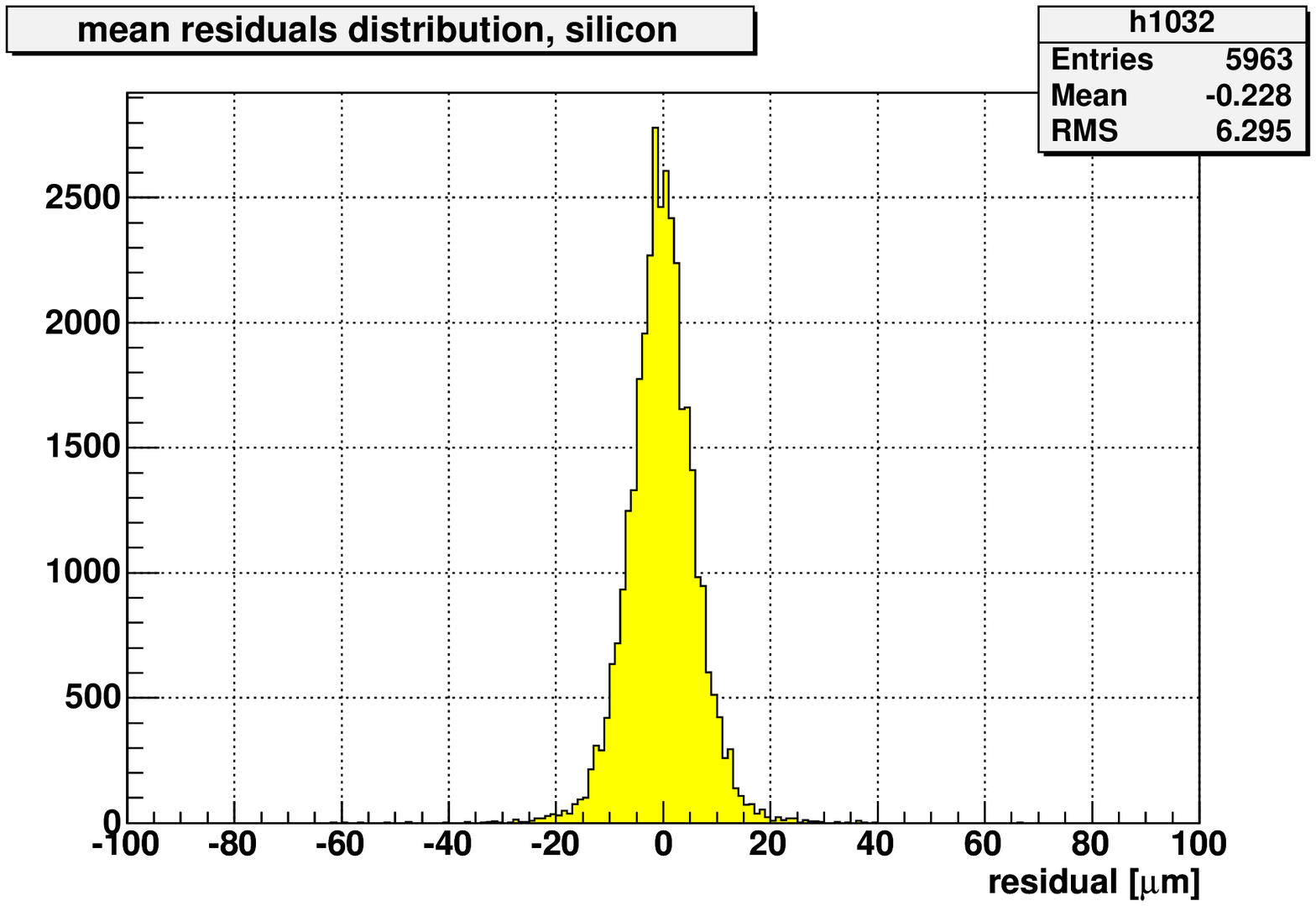}
\includegraphics[width=10cm,height=6cm]{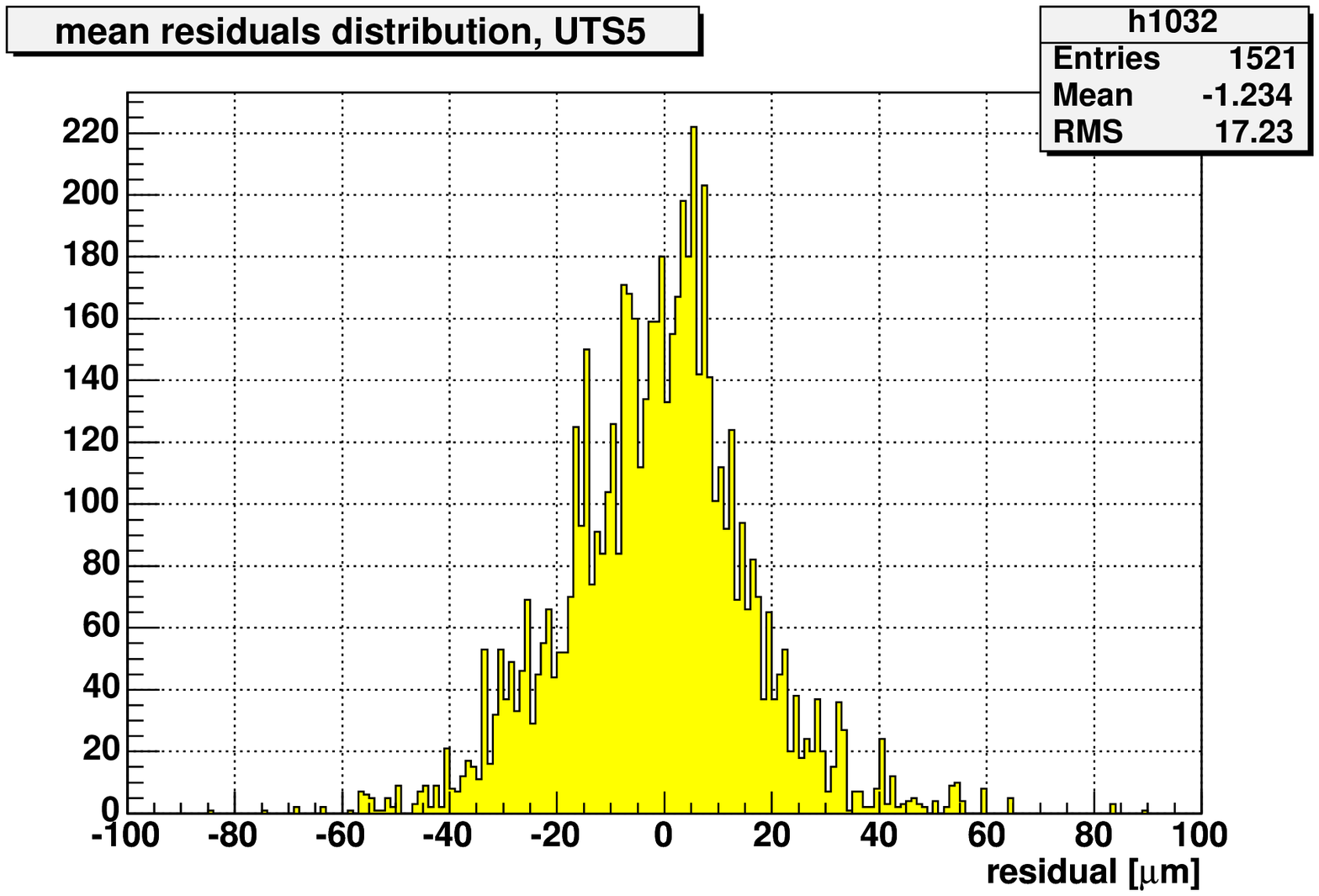}
\includegraphics[width=10cm,height=6cm]{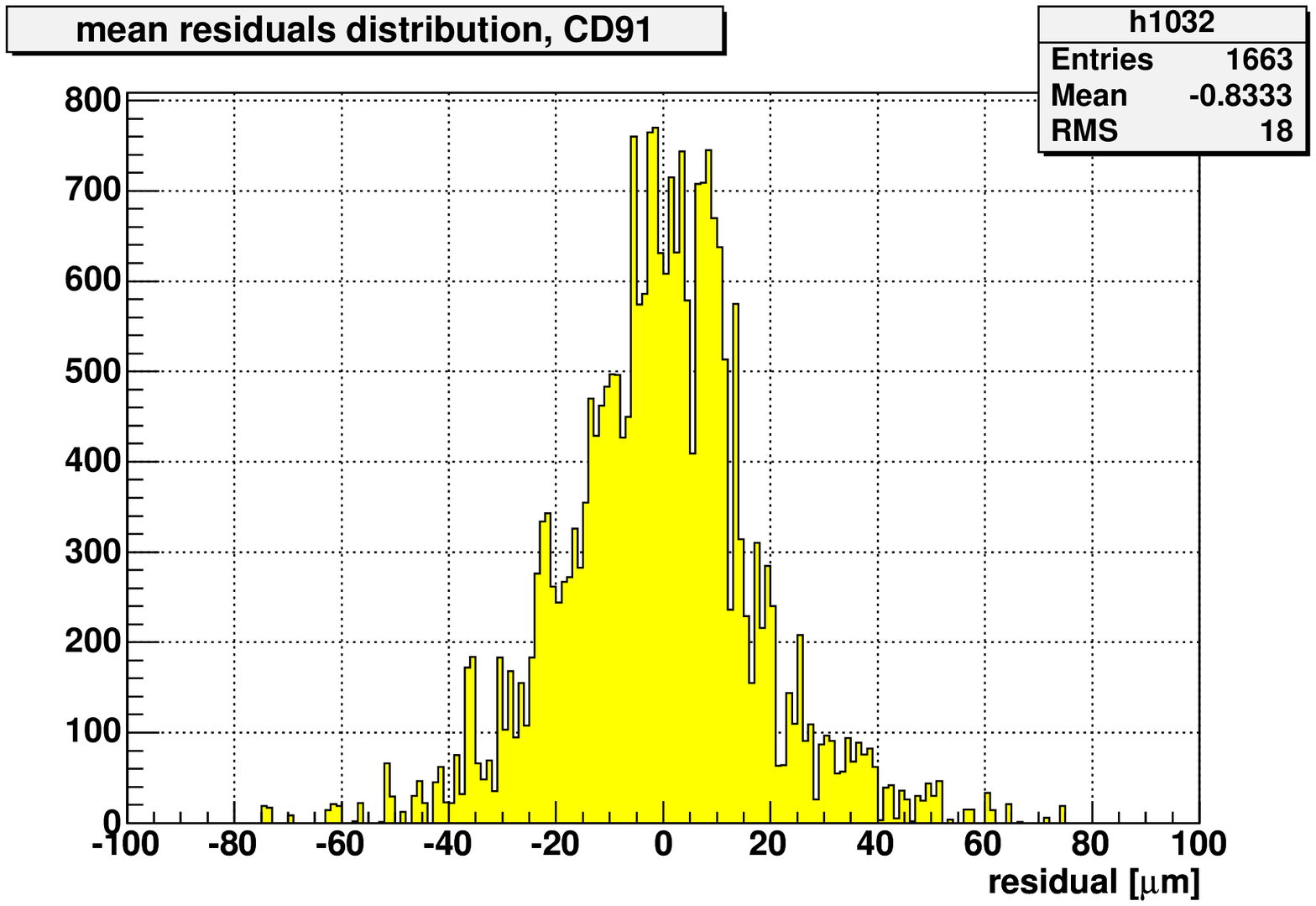}
\end{center}
\caption{\label{MeanRes} Distribution of the 
mean spatial residuals shown in Fig.~\ref{ResidualsMap}.  
Each entry (mean spatial residual at a given position) 
is weighed with the number of tracks
used in the computation of the mean residual.  
The upper plot is for the silicon sensor, the
middle one for the diamond sensor UTS-5 and the lower plot is for 
the diamond sensor CD91.} 
\end{figure}

The statistical error $\delta r_i$ 
on each mean residual $r_i$ is $\sigma/\sqrt{n_i}$ where $\sigma$ is the 
spatial resolution and $n_i$ is the number of events inside the position bin.
The weighted r.m.s of mean residuals is  
$R = \sqrt{ \Sigma_i (r_i/\delta r_i)^2/\Sigma_i 1/\delta r_i^2} 
= \sqrt{ \Sigma_i n_i r_i^2/\Sigma_i n_i}$.   
This is $(17.3 \pm 0.3) \;\mu$m for UTS-5,  
$(18.0 \pm 0.3) \; \mu$m for CD91 
and $(6.46 \pm 0.05) \;\mu$m 
for the silicon sensor. 
The weighed distribution of the average residuals is shown 
in Fig.~\ref{MeanRes}.

In absence of systematic effects on the spatial response of the detector 
one expects the r.m.s. of mean residuals R 
to be $R_{\mbox{stat}} = \sigma/\sqrt{N}$ 
where $N$ is the average number of entries for each position bin. 
For silicon it is $R_{\mbox{stat}} = (6.27 \pm 0.04) \;\mu$m which is indeed 
close to the measured value. For diamond an important systematic 
contribution exists, since the measured values of $R$ are much larger 
than the values expected from statistic fluctuations, 
$R_{\mbox{stat}} = (11.65 \pm 0.18) \;\mu$m for UTS5 and
$R_{\mbox{stat}} =  (6.53 \pm 0.08) \;\mu$m for CD91. 

To quantify the apparent clustering of residual shifts,
the linear correlation coefficient between the residuals of all 
track pairs has been determined. Since each event which passes 
selection cuts has only one reconstructed track, the pairs are 
formed considering the tracks of two different events.
Correlation coefficients are 
determined in bins of the distance between the two tracks 
of a pair. An empirical fit function 

\begin{equation}
\label{CorrFit}
A\exp(-x/x_0) + a \sin(x/50 \mu\mbox{m} + b)/\sqrt{x}
\end{equation}

has been found to describe the data of the correlation 
coefficient of residuals as a function of track separation $x$.
The function is the sum of two components. A sinusoidal 
modulation with the periodicity of the pixel pitch 
describes the correlation due to the pattern structure 
of the sensor. A falling exponential accounts for the 
residual shifts introduced by the grain structure. 
The denominator in the exponent $x_0$ we term the   
correlation length. It is a measure of the scale 
of the cluster structure of similar residual shifts.
 
\begin{figure}[!ht]
\begin{center}
\includegraphics[width=10cm]{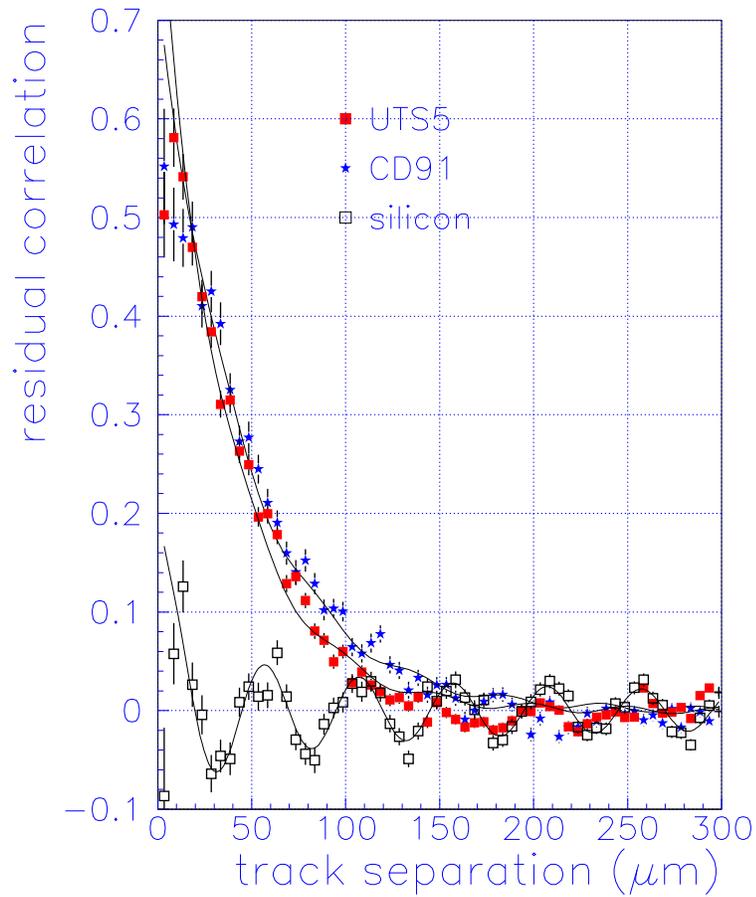}
\end{center}
\caption{\label{Autocorrelation} Correlation between the spatial 
residuals of two events, as a function of the separation between the 
tracks of the two events.}
\end{figure}

The residuals correlation is reported in Fig.~\ref{Autocorrelation}. 
For silicon the correlation is well described by the 
pitch modulation term only and the amplitude of 
the exponential term is consistent with 
zero. 

For diamond the exponential peak is evident. The 
fit gives a correlation length $x_0 = (36.0 \pm 0.5)\mu$m for UTS5 
and $(44.4 \pm 0.8 )\mu$m for CD91. The correlation amplitude $A$ is 
$0.826 \pm 0.016$ for UTS5 and $0.748 \pm 0.018$ for CD91.
 
The presence of the exponential peak in the diamond correlation 
distribution is 
evidence of non-homogeneous charge collection properties producing 
local systematic shifts in the 
position response. The investigation of their origin is the subject of 
the studies described in the next chapter. 

\section{Simulation of the diamond sensors} 

\label{sec3}

In order to understand the experimental results in terms of the 
microscopic properties of charge carriers drift in diamond, a
numerical simulation of the response of diamond sensors to ionizing 
radiation has been developed.

\subsection{The model}
\label{model}

The interactions of high-energy pions with the diamond sensor have 
been simulated using the GEANT4 package~\cite{GEANT4}. This step of 
the simulation produces a file with a list of the energy deposits of the 
particles inside diamond (hits). The energy of each hit is converted 
into a number of electron-hole pairs using the conversion factor of 
one pair for 13.1~eV of deposited energy. The charge carriers are 
then drifted in the diamond in steps of $l=5 \; \mu$m until they are 
trapped or they are collected at the electrodes. 

{\bf Drift model:}
The drift direction is determined by the local electric field and thermal 
diffusion. The local electric field is the superposition of the external 
field of 1040~V/mm and the polarization field created by the charge 
carriers trapped in the sensor. The diffusion function is a Gaussian with  
$\sigma =  \sqrt{2Dt}$, where $D$ is the diffusivity and $t$ the drift 
time. The diffusivity $D$ is related to the low-field mobility $\mu_0$ 
and the temperature $T$ by the Einstein relation $D = \mu_0 kT/e$, 
where $k$ is the Boltzmann constant and $e$ the elementary charge.

The drift time corresponding to one drift step is $t=l/v$ where the 
drift velocity $v$ depends on the electric field. It is proportional to 
the electric field for low field intensity, and reaches saturation in the 
high-field limit. The following parametrization was used to connect the 
two asymptotic behaviors~\cite{Jac77}

\begin{equation}
v(E) = \frac{\mu_0 E}{\left[ 1+\left(\frac{\mu_0 E}{v_s}\right)\right]} 
\end{equation}

where $\mu_0$ is the low-field drift mobility and $v_s$ is the 
high-field saturation velocity. The values of these parameters in 
diamond are controversial~\cite{Pan93}. The following values 
have been used~\cite{Can79,Art79}: 
$\mu_0 = 2400 \;\; \mbox{V}^{-1} \mbox{cm}^2\mbox{s}^{-1}$, 
$v_s = 1.5 \cdot 10^{7}$~cm/s for electrons and 
$\mu_0 = 2100 \;\; \mbox{V}^{-1} \mbox{cm}^2\mbox{s}^{-1}$, 
$v_s = 1.05 \cdot 10^{7}$~cm/s for holes. 

\begin{figure}[!ht]
\begin{center}
\includegraphics[width=10cm,height=8cm]{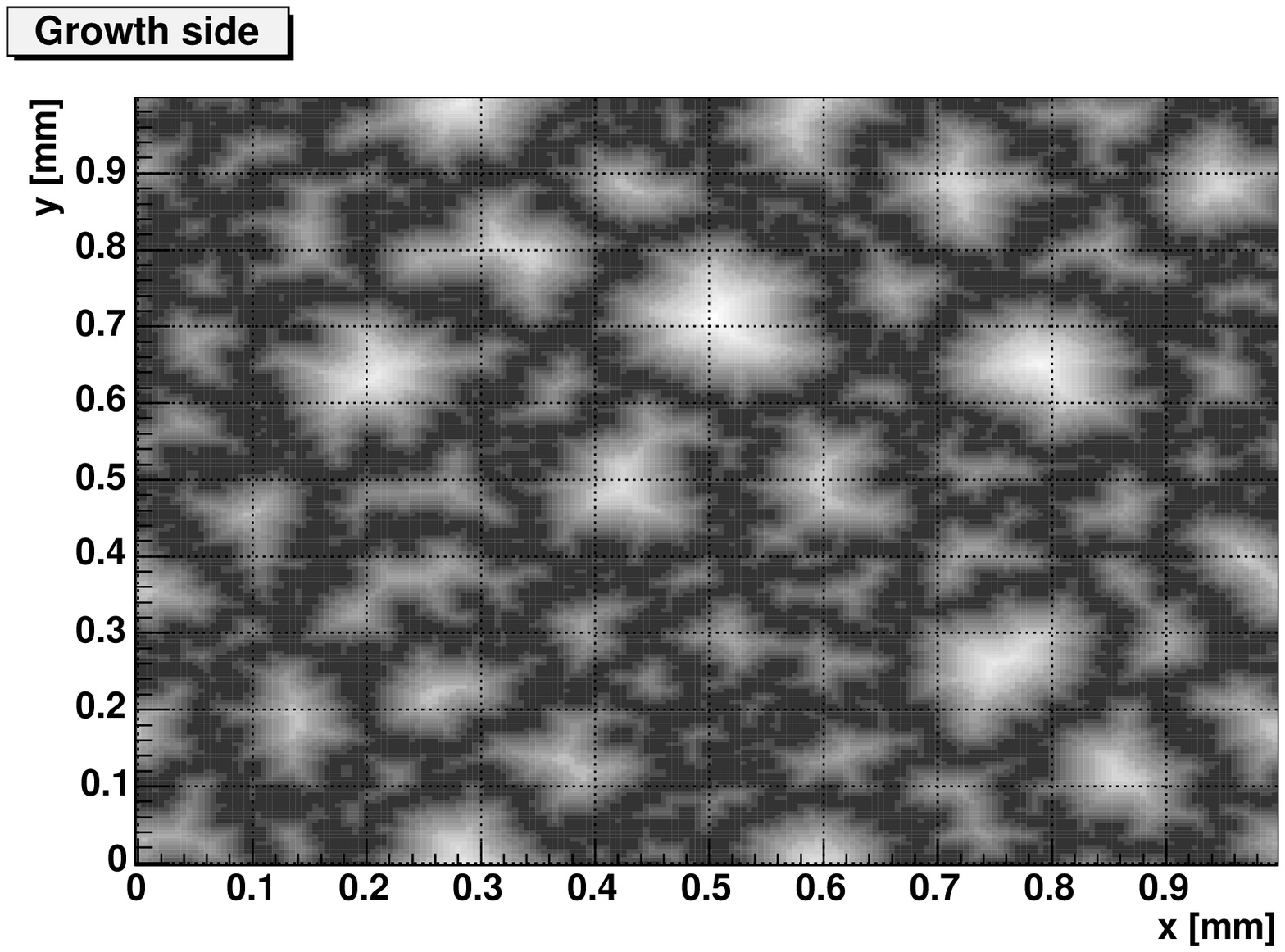}
\includegraphics[width=10cm,height=8cm]{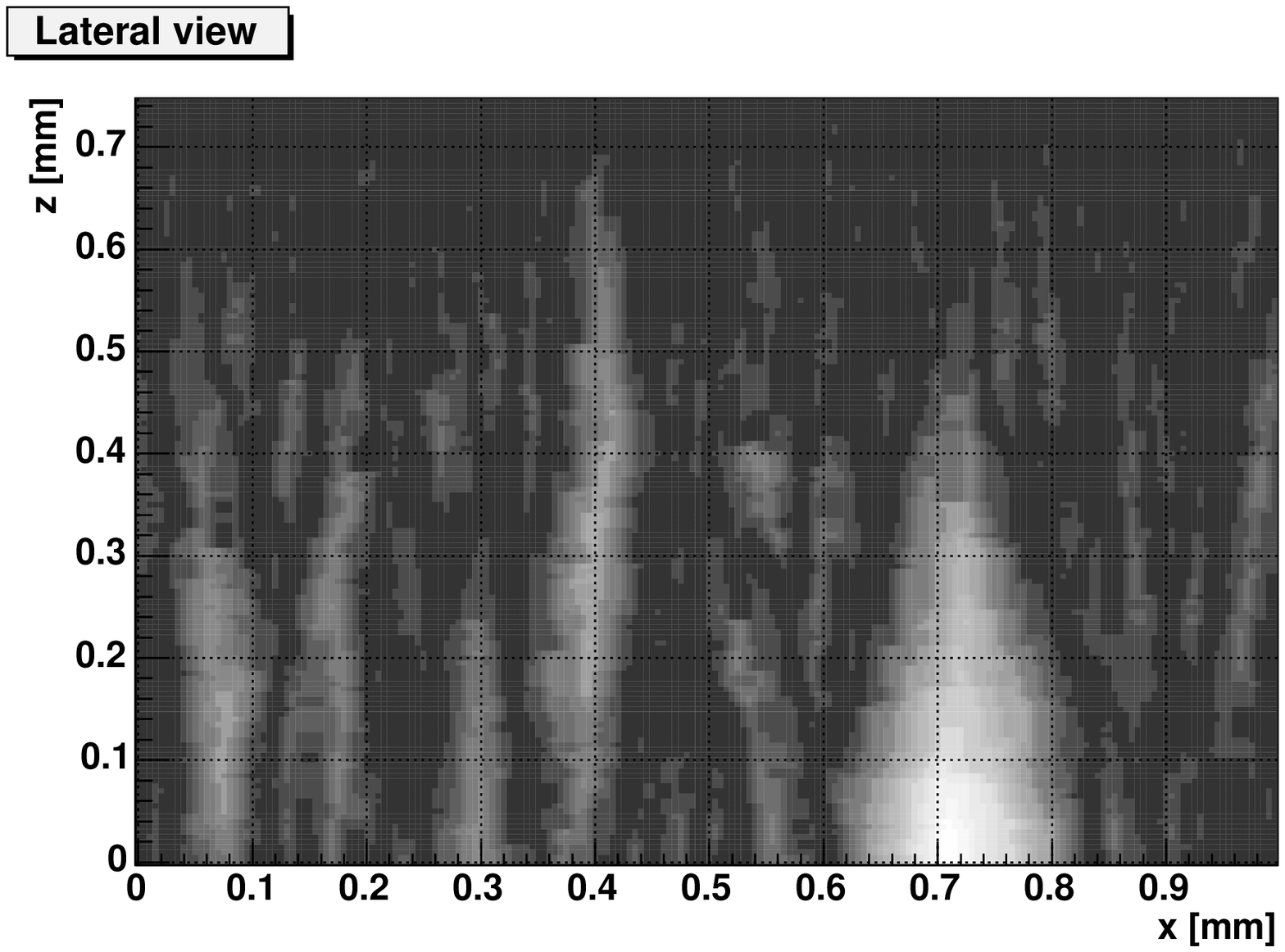}
\end{center}
\caption{\label{GrainMap} Three-dimensional 
simulation of the growth of diamond crystals. The grey scale   
is proportional to the carrier lifetime, so that the crystallite boundaries 
are visible as dark region, and the crystallites bulk is white.
The upper plot shows the crystallite shapes in a plane perpendicular 
to the growth direction, on the growth side of 
the diamond film (where the pixel implantations are 
made). The lower plot shows the same in a lateral view; the average 
dimension of crystallites increases from the substrate side of the film 
(below) to the growth side (above). A thickness of 0.3~mm on the 
substrate side has been removed after film growth and it is not 
used in the simulation either.} 
\end{figure}

{\bf Diamond grains 
generator:} The trapping model presented below requires the 
simulation of the grain structure of CVD diamond.  
The grains growth generator is a generalization to three dimensions 
of the generator described in~\cite{Oh99}. 
The space is divided in small cubic cells 
($5 \times 5 \times 5 \; \mu\mbox{m}^3$) 
and the grains are initially 2 x 2 cells wide.  
Every time that a layer of cells is added to the diamond film, each 
cell near the grain border  
has a probability $p$ to be claimed by a neighboring grain. 
The resulting conflicts between 
grains claiming the same space element are solved assigning to  
each grain a probability to defeat the other proportional to the 
grain width. This results in large grains getting larger and small grains 
getting smaller and finally being overgrown.
The resulting simulated structure is shown in Fig.~\ref{GrainMap}. 

The only parameters of the model are the growth probability $p$ and the 
film thickness, which is the sensor thickness plus the material 
removed from the substrate side.
The film thickness (732~$\mu$m) is equal to the final sensor thickness plus 
the thickness of material removed from the substrate side.
The growth probability is chosen so that the resulting 
spatial residuals correlation length (see Sec.~\ref{results}) 
agrees with the data.

{\bf Trapping model:}
The probability for a charge carrier to be trapped within the drift 
step $l$ is $1-\exp(-l/v\tau)$ where $v$ is the velocity and $\tau$ 
the lifetime.

The lifetime is inversely proportional to the local density of active 
traps. Different initial trap density distributions have been 
implemented. 
In the simulation presented in this paper, the initial trap density   
depends on the shortest distance to the 
next grain boundary and is described by~\cite{Oh99} 

\begin{equation}
\label{eqTrap}
n = n_0/[1-\exp(-r/r_0)]
\end{equation}

with $r$ the distance to the next grain boundary, 
$n_0$ the minimum 
trap density (deep in the grains bulk) and  
$r_0$ the {\em lifetime length}. The trap density becomes very high 
near the grain boundaries. This model is supported by observations that
the impurity concentration is strongly enhanced at grain 
boundaries~\cite{Pol96,Oh98} thus resulting in a larger density of 
electrically active traps~\cite{Oh99,Pol99}.

Priming is simulated by implementing two classes of traps~\cite{Oh99}. 
After they capture a charge, the traps of one class 
can act as recombination centers, while the traps of the other  
class are permanently passivated.     
The electron and hole lifetimes are computed as 

\begin{equation}
\tau_e = 1/\beta_e (n_{0} + n_{+})
\end{equation}

\begin{equation}
\tau_h = 1/\beta_h (n_{0} + n_{-})
\end{equation}

$n_{0}$ is the density of 
unfilled traps of both classes, $n_{+}(n_{-})$ is the density of 
recombination centers  
filled with holes (electrons). $\beta$ can be interpreted
as the product of the charge carriers thermal velocity and an effective
trapping cross section. It is the same for traps of the two classes but 
different values can be set for electrons and holes.

When a charged carrier is trapped, the local density 
of filled and unfilled traps  and the local trapped charge density 
are updated. The polarization field created 
by the trapped charges is periodically updated 
and superimposed to the external 
field to get the electric field map.    

As the diamond sensor is exposed to ionizing radiation, three 
dose-dependent effects occur:

\begin{itemize}

\item The pure trapping centers are passivated, so that the effective 
trap density is reduced, simulating the priming effect.    

\item Near the electrodes  
the charge carrier which is drifting toward the electrode is dominant 
(electrons near the pixels electrodes, holes on the other side). Thus 
a net space charge density develops. The resulting polarization  
screens the external field. 

\item A net space charge develops also near the grain boundaries
(only in the scenario with non-uniform trap density).
Because of grain lateral growth the boundary surface makes 
in general a non-zero angle with the $z$ direction. Electrons and 
holes approach this surface from opposite directions, and 
are trapped before reaching the interface because of the very 
high trapping density.  Hence a space charge distribution 
is created, with opposite signs  on the two sides of the interface.
The electric field they produce  is normal to the interface, and 
has a component which is transverse to the external field. 
This is the source of the charge drift parallel to the sensor surface. 

\end{itemize}

\begin{figure}[!ht]
\begin{center}
\includegraphics[width=10cm]{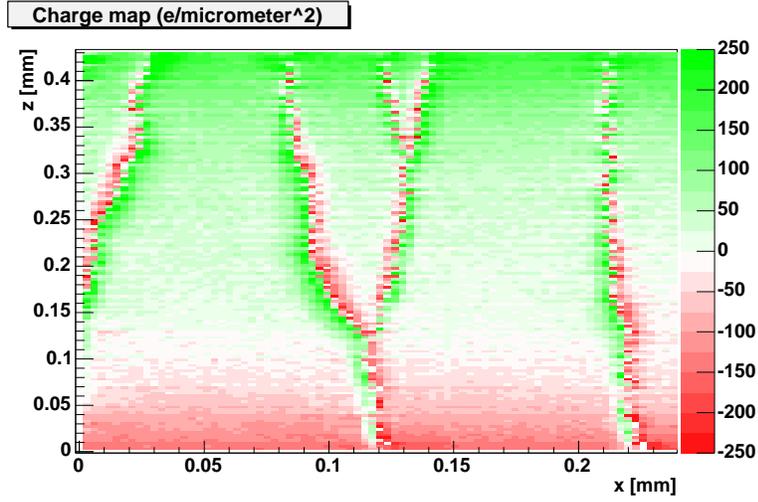}
\end{center}
\caption{\label{Polarization} Trapped charge as a 
function of position inside the sensor, for a two-dimensional simulation 
with preferential trapping near the grain boundaries.  
The crystal growth direction and the external electric field are along 
the $z$ direction;  the pixel implants are located at $z=0$.
The electrons move downward. The grain boundaries are visible as 
tree-like structures. The pixel electrodes are at the bottom.}
\end{figure}

The distribution of trapped charge for   
a two-dimensional simulation is shown in  Fig.~\ref{Polarization}.  
A net negative trapped charge develops near the pixels at the bottom 
and a net positive charge near the other electrode. The 
tree-like structures are the boundaries between grains (growing 
from above). There is a larger trapped charge there, negative 
above the boundary and positive below it.

The parameters of the trapping model discussed are  
$n_0, r_0, \beta_e, \beta_h$ and the fraction $\alpha$ of pure 
trapping centers as explained above. A constraint on the 
above parameters is provided by 
the measured value of the average collected charge (3000~$e^-$). 
This still leaves four degrees of freedom in the parameter choice. 
Since the other parameters are related to unmeasured properties 
(such as the priming ratio, or the hole/electron lifetime ratio)
it is not possible to determine them from the data.

Some educated guesses have been made instead, with the purpose 
of building a model which can qualitatively reproduce the features 
observed 
in the data, in particular the systematic position shifts.
The choice of final values for the parameters is reported below. 

\begin{figure}[!ht]
\begin{center}
\includegraphics[width=10cm]{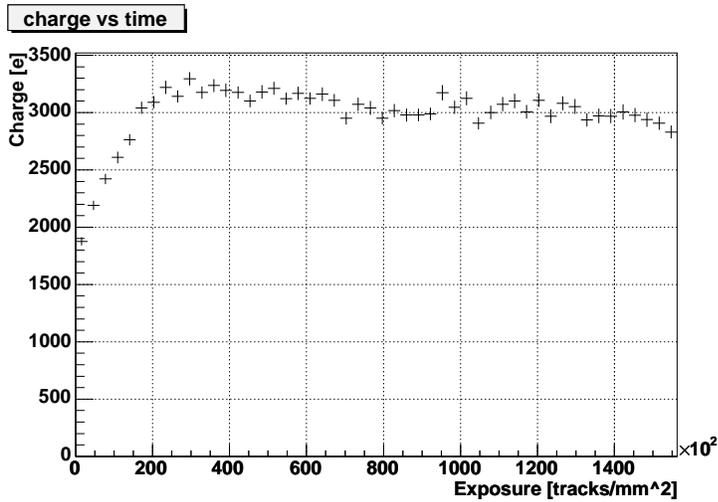}
\end{center}
\caption{\label{Priming} Priming curve: average charge collected as a  
function of the exposure resulting from the detector simulation. }
\end{figure}

\begin{itemize}

\item The ratio between 
hole and electron trapping $\beta_h/\beta_e$ was chosen 
to be 1.5 since measurements with $\alpha$ particles show 
that holes have a lower lifetime 
than electrons in diamond sensors~\cite{Beh98}. 

\item The fraction $\alpha$ of traps which are permanently 
passivated upon capture of a charge carrier affects primarily the 
priming ratio (the ratio 
between the signals observed in the primed and unprimed state).
The value $\alpha = 0.85$ was used. The resulting priming ratio 
is 1.6 (it is also affected by $n_0$, discussed below). 
The priming ratio of the sensor studied at the test beam is not known, 
but the simulated value is typical for sensors with good charge 
collection efficiency~\cite{Oh98}. 
The simulated priming curve is shown in Fig.~\ref{Priming}.

\item The extent of the region near the 
crystallite boundaries where trapping is enhanced 
is determined by $r_0$. 
The value $r = 25 \; \mu$m was chosen. This is smaller than the 
crystallite size so that in the crystallite bulk the 
charge lifetime is indeed larger than near the border. 
A lower value does not change qualitatively the results 
presented in the next section, but would make them more 
sensitive to the cell length $l$ of the grid of points on 
which the trap densities are computed. 

\item The total trap concentration $n_0$ at the center of   
crystallites. The value $n_0 = 2.5 \; \mu\mbox{m}^{-3}$ 
has been used. 
A smaller value results in a weaker polarization field and thus 
in a smaller lateral component of the charge drift, in contrast 
with the experimental evidence. A larger trap density results   
in too effective external field screening, 
making it difficult to reproduce a typical priming curve 
(the signal reaches a maximum then decreases with dose). 

\end{itemize}

The sensitivity of the results of the simulation on each parameter 
has been computed varying one parameter at a time and retuning 
$\beta$ (keeping constant $\beta_e/\beta_h$) so that the 
mean signal was always equal the experimental value. The 
resulting variations of efficiency, resolution and systematic shift 
r.m.s. $R$ are reported in table~\ref{sens}.

{\bf Signal formation:}
At the end of the charge drift algorithm, a list of the positions where 
the electrons and holes have been trapped or collected is produced. 
The charge $Q$ induced on a given pixel electrode is computed as 

\begin{equation}
Q = \Sigma_i q_i[\Phi_w(\vec r_i) - \Phi_w(\vec r_0)]
\end{equation}

where $q_i$ are the drifting electric charges, 
$\vec r_i$ is the position at which they were   
trapped or collected, $\vec r_0$ their initial positions, 
and $\Phi_w$ is the 
weighting potential~\cite{Ram39,Rad88} for the pixel electrode under 
consideration.

The experimental electronics threshold and noise (1000~$e^-$ and 
200~$e^-$ r.m.s.) are simulated.

The pixel hits are processed with the same algorithms used 
for the analysis of real data. To simulate the telescope extrapolation 
uncertainty of 6~$\mu$m a Gaussian smearing is added to the 
true track position. 

\subsection{Results}

\label{results}
Due to the severe CPU and memory requirements of a 3D simulation, 
only a $1 \times 1 \times 0.45 \; \mbox{mm}^3$ volume 
was simulated.
Trap density maps and charge drift were made with a discretization
of space in 
$5 \; \mu$m steps while the computation of the polarization
field was made with a grid with $20 \; \mu$m pitch. 

The diamond samples were in a primed state when data were collected. 
This was simulated with a sufficiently large number of traversing 
charged particles  to reach a primed state before events for analysis 
were processed.
 
\begin{figure}[!ht]
\begin{center}
\includegraphics[width=10cm]{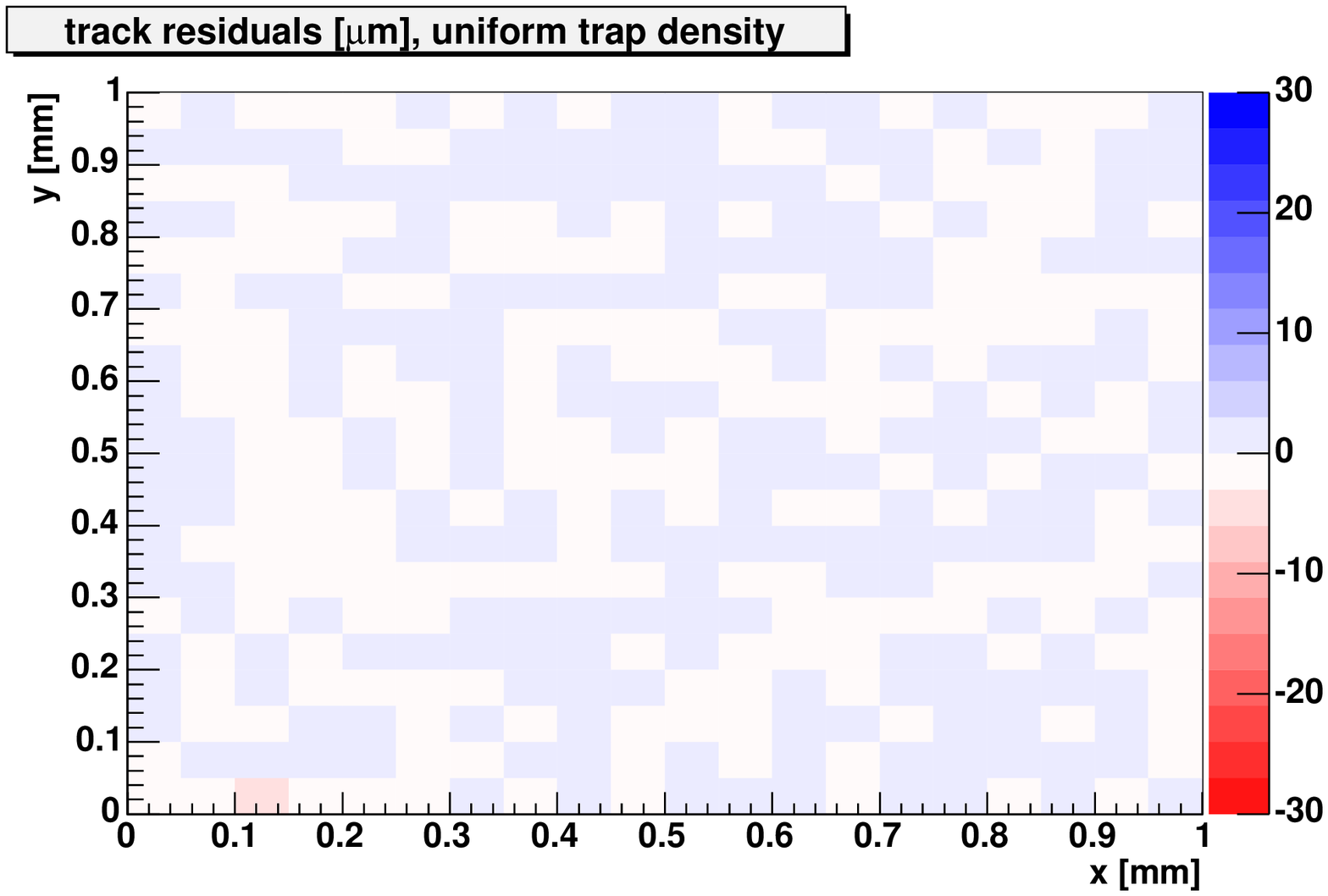}
\includegraphics[width=10cm]{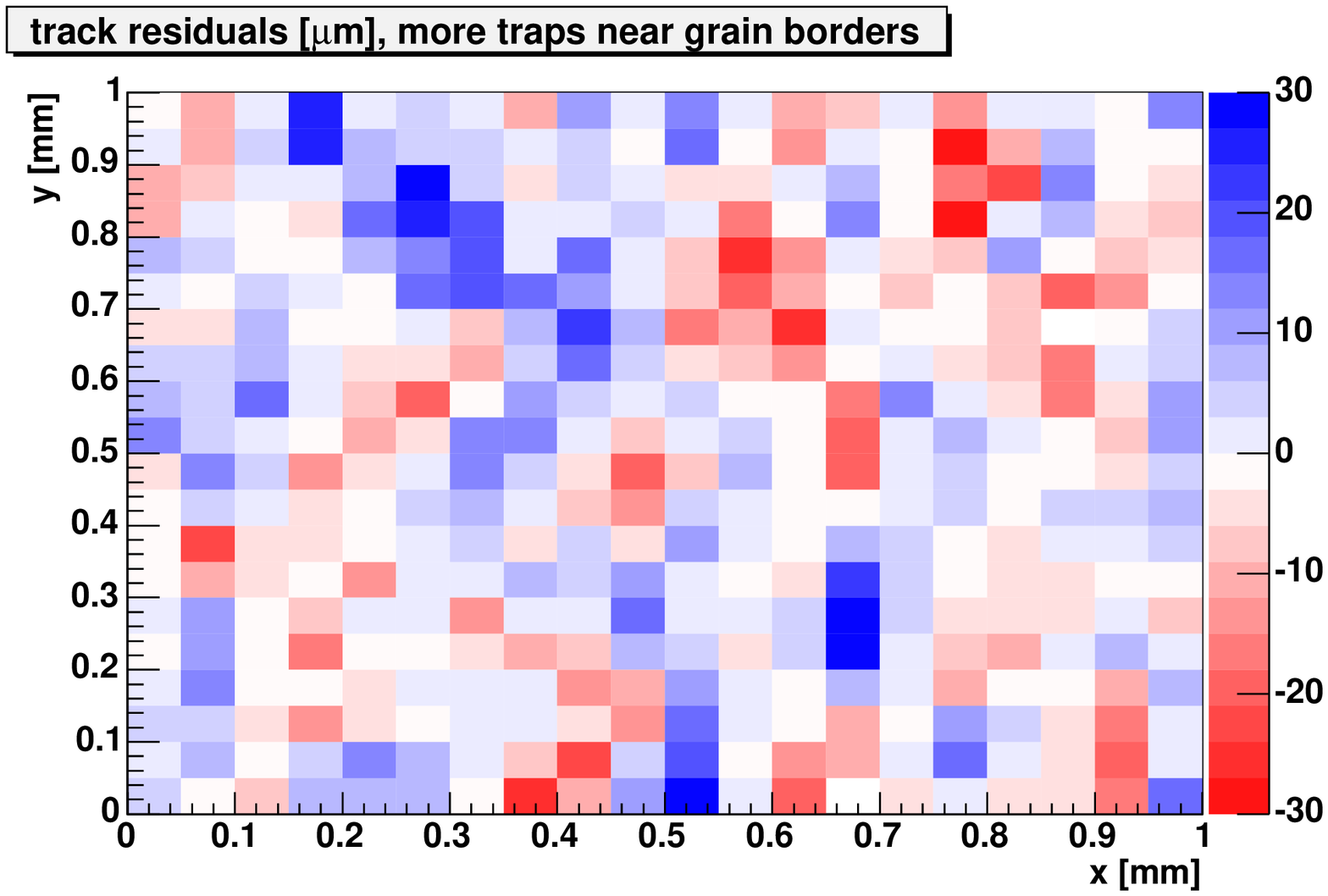}
\end{center}
\caption{Mean spatial residual as a function of track position.
The upper plot is obtained with the simulation with uniform  
trap centers density, the lower plot using a higher trap density near the 
grain boundaries as described in the text.
\label{simresmaps} }
\end{figure}

The simulated mean spatial residuals as a function of track 
position are shown in the lower plot of   
Fig.~\ref{simresmaps}. For comparison, the upper plot
reports the results obtained with a uniform 
trap density (i.e., without simulating the crystallites 
structure). 

The simulation with trapping enhanced near the crystallites boundaries
shows systematic shifts as large as 30~$\mu$m in the position 
of the collected charge, as observed  
with the data (Fig.~\ref{ResidualsMap}). The 
agreement with experimental
results is much better than using the simulation with  
uniform trapping\footnote{The simulation with uniform trapping has an r.m.s.
of mean residuals which is even smaller than the value measured with 
the data of the silicon sensor. This is due to the fact that the 
average number of events per bin is larger in the simulation, so 
statistical fluctuations are smaller.}.

\begin{figure}[!ht]
\begin{center}
\includegraphics[width=10cm]{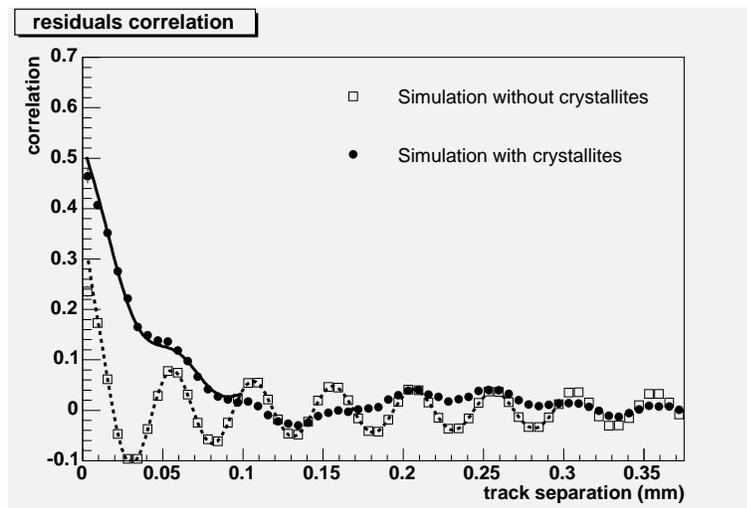}
\end{center}
\caption{Track residuals correlation as a function of track position.
The squares, fitted with a dashed line, are obtained with the simulation 
with uniform  trap centers density. 
The circles, fitted with a solid line, are obtained using a higher trap 
density near the 
grain boundaries as described in the text.
\label{simcorr} }
\end{figure}

The same conclusion holds for the residual correlation 
plot (Fig.~\ref{simcorr}). In the simulation with uniform trapping 
(square points) only the readout pitch correlation
is observed, and a good fit is obtained omitting the exponential 
term of eq.~\ref{CorrFit}, as with the experimental silicon data. 
For the simulation with non-uniform trapping, however,  
the correlation at small distance scales 
seen in the diamond data (Fig.\ref{Autocorrelation}) is reproduced. 
The simulated residual correlation plot has been fitted with 
eq.~\ref{CorrFit}. 

The correlation length is determined by the average crystallite size, 
determined by the growth parameter $p$, and by the trapping parameters, 
especially the spatial scale for trapping non-uniformities $r_0$. 
The correlation length was tuned to the experimentally measured 
value by varying $p$ while keeping the trapping parameters constant,
except for an overall lifetime scaling to get an 
average signal equal to the measured value.
As $p$ is increased the average crystal size\footnote{The average
was computed assigning to each crystallite cross section 
$a_j$ a weight proportional to the probability of a track to cross it.
Thus, $A = \Sigma a_j^2 / \Sigma a_j$.}
at the growth surface increases from  
0.0119~$\mbox{mm}^2$ to 0.0422~$\mbox{mm}^2$ and the correlation 
length increases from $(21.7 \pm 0.8) \; \mu$m to 
$(70.4 \pm 1.0) \; \mu$m. 
The final choice of $p=0.15$, corresponding to an average 
grain size $A = 0.0238~\mbox{mm}^2$, gives a correlation 
length of $(36.3 \pm 0.4) \;\mu$m, in agreement with the observed values.

In table~\ref{SimTab} the values of the spatial resolution, the 
r.m.s. of systematic shifts R (computed as 
in section~\ref{analysis}), the
detection efficiency and the residuals correlation length 
are reported for the two simulations and test-beam data.
The statistic error is quoted for simulated values. 

The systematic shifts are absent in the simulation without the 
crystallite structure 
since they are produced by the lateral component of the 
polarization field created by charges trapped near the 
crystallite boundaries. The simulation of the polycrystalline    
structure is also needed to reproduce the 
experimental observed detection efficiency of about 70\%. When it 
is omitted the efficiency (respecting the experimental 
constraints on the 
average cluster charge and discriminator threshold) is 
almost 100\%. The simulation of the crystallites 
results in a non-uniform charge collection and regions 
with reduced detection efficiency. This reproduces 
the experimentally observed detection efficiency to 
reasonable accuracy.

While the simulation with enhanced trapping near the grain borders 
is able to qualitatively reproduce the lateral displacement of charge 
collection and the residuals correlation at small distance scales 
observed in the data, these polarization effects are less pronounced 
in the simulation. 
As a result, the amplitude and the r.m.s. of systematic shifts $R$ 
are smaller and the spatial resolution is better in the simulation 
than in the data.

It is possible to increase the strength of polarization effects with 
a different choice of the model parameters. In particular, a larger 
value for $n_0$ increases the density of traps and thus the polarization 
charges. However, this also makes the screening of the external electric 
field more effective, and it is no longer possible to obtain a priming 
curve with a flat plateau (Fig.~\ref{Priming}), instead the signal reaches a 
maximum and then decreases, as the field screening prevails over the 
passivation effect. We have restricted ourselves to choices of the 
model parameters which give a priming curve typical for diamond 
sensors exposed to minimum ionizing particles. 

Most likely, to achieve a quantitative agreement with the data requires a 
more accurate trapping model. In the model described in 
Section~\ref{model} only two classes of traps have been implemented, 
pure trapping centres with cross sections $\sigma_e$ and $\sigma_h$ 
for electrons and holes respectively and trapping/recombination 
centres with the same 
cross sections $\sigma_e$ and $\sigma_h$ both for trapping and 
recombination processes. In general, several classes 
of trapping centres should be anticipated, each one with four different 
cross sections
for electron trapping, hole trapping, electron recombination and 
hole recombination, respectively. 
In order not to spoil the predictive power of 
the model by a too large number of free parameters we refrained from 
introducing these parameters. A
detailed microscopic characterization of the number, 
spatial distribution and trapping cross sections of defect centres 
in the diamond sensor would be required to better assess the role of
these trapping centres and include them in the simulation. 
That is, however, beyond the scope 
of the study presented in this paper.

\section{Conclusions}

Test-beam data were taken with diamond pixel detectors. A detection
efficiency of about 70\% and a spatial resolution of about 
20~$\mu$m were obtained. 
The fine segmentation provided by the pixels and the tracking of 
the particles by a silicon microstrip telescope allowed a study of 
the spatial resolution of the detector as a function of the incident 
position of the particles. 

The spatial resolution was found to be degraded
by the presence of regions  
where the reconstructed impact position given by the 
pixel cluster and the true track position measured with a beam 
telescope deviate systematically. These regions have  
the same dimension of the diamond crystallites (about 100~$\mu$m) 
hence they appear to be related to the polycrystalline nature 
of the sensor.

A model was proposed which attributes the charge collection
position shifts to the polarization field created by charges
trapped near the crystallite borders, which can have a 
component parallel to the sensor surface. 

A simulation was presented which implements 
a detailed description of the ionizing 
particle interactions, the drift and the trapping of charge 
carriers in diamond, polarization effects and signal induction 
on the electrodes. 
A model for CVD diamond growth was deployed to simulate the 
crystallite shape in three dimensions, together with a model for the 
trap density distribution in crystallites.

The simulation is able to reproduce at least qualitatively the lateral 
displacement of charge collection and the residuals correlation at 
small distance scales observed in the data, albeit the 
effect is weaker in the simulation. Detection efficiency for  
the same amount of collected charge is also reproduced.
With a uniform trap density  
neither the charge collection displacements and residuals 
correlation nor the detection efficiency can be reproduced.

{\bf Acknowledgments}

The diamond sensors of the pixel detectors used for the test beam 
studies were provided by the RD42 collaboration. The front-end 
electronics chips were provided by the ATLAS Pixel collaboration, 
which we also acknowledge for use of the test beam setup and beam 
time. The decoding and reconstruction of test beam data was 
made with a program developed by A. Andreazza, C. Meroni, F. Ragusa
and C. Troncon (INFN and University of Milan).

\begin{table}[!p]
\begin{center}
\begin{tabular}{||c|c|c|c||  }
\hline
sensor                    &UTS-5 & CD91 & silicon \\
\hline
detection efficiency (\%)  &  $76.8 \pm 0.5$   & $67.69 \pm 0.10$  & 
                              $99.61 \pm 0.03$   \\
mean cluster size          & $1.484 \pm 0.010$ & $1.260 \pm 0.001$ 
                           & $1.301 \pm 0.003$   \\
spatial resolution ($\mu$m)& $23.35 \pm 0.21$  & $25.45 \pm 0.11$  
                           & $14.46 \pm 0.05$ \\
 $R_{\mbox{meas}}$ ($\mu$m)& $17.3 \pm 0.3$ & $18.0 \pm 0.3$ 
                           & $6.46 \pm 0.05$ \\
 $R_{\mbox{stat}}$ ($\mu$m) & $11.65 \pm 0.18$ & $6.53 \pm 0.08$ 
                           & $6.27 \pm 0.04$  \\ 
correlation length ($\mu$m)& $36.0 \pm 0.5$   & $44.4 \pm 0.8$ & 0   \\
correlation amplitude      & $0.826 \pm 0.016$ & $0.748 \pm 0.018$ & 0 \\
\hline                         
\end{tabular}
\end{center}
\caption{\label{DataTab} Summary of the measurements performed on the 
two diamond detectors. The values for a silicon detector with the same 
geometry and electronics is presented for comparison. } 
\end{table}

\begin{table}[!p]
\begin{center}
\begin{tabular}{||c|c|c|c||  }
\hline
parameter variation        & $\Delta \epsilon$(\%) 
                           & $\Delta \sigma$ ($\mu$m)  
                           & $\Delta R$ ($\mu$m) \\
\hline
$\beta_h/\beta_e \pm 0.5$  & $\pm 0.5$ & $\pm 0.6$ & $\pm 0.7$ \\
$r \pm 5 \mu$m             & $\pm 2$   & $\pm 0.2$ & $\pm 0.1$  \\
$\alpha \pm 0.05$          & $\pm 3$   & $\pm 0.8$ & $\pm 1.1$ \\  
$n_0 \pm 0.5 \; 
\mu\mbox{m}^{-3}  $        & $\pm 2$   & $\pm 0.8$ & $\pm 0.8$ \\
\hline                         
\end{tabular}
\end{center}
\caption{\label{sens} Sensitivity of the simulation results on parameter 
choice.} 
\end{table}

\begin{table}[!t]
\begin{center}
\begin{tabular}{||c|c|c|c|c||  }
\hline
sensor                   &UTS-5 & CD91 & sim. I & sim. II \\
\hline
 spatial resolution ($\mu$m) 
  & $23.35 \pm 0.21$ & $25.45 \pm 0.11$& $19.22 \pm 0.06$& $15.12 \pm 0.04$ \\
 $R$ ($\mu$m) 
  & $17.3 \pm 0.3$ & $18.0 \pm 0.3$ & $8.8 \pm 0.3$ & $1.08 \pm 0.04$ \\ 
 detection efficiency (\%)
  & $76.8 \pm 0.5$ & $67.69 \pm 0.10$ & $58.10 \pm 0.18$ & $99.54 \pm 0.02$ \\
 correlation length ($\mu$m)    
  & $36.0 \pm 0.5$ & $44.4 \pm 0.8$ & $36.2 \pm 0.5$ & 0  \\
 correlation amplitude & $0.826 \pm 0.016$ & $0.748 \pm 0.018$ 
                       & $0.492 \pm 0.007$ &  0 \\ 
\hline                         
\end{tabular}
\end{center}
\caption{\label{SimTab} Spatial resolution, r.m.s. of systematic 
shifts $R$ (defined in section~\ref{analysis}), 
efficiency, residuals correlation length and amplitude for the data 
taken with the 
UTS-5 sensor, the data taken with the CD91 sensor, 
the simulation with trapping enhanced near the grain borders (I) 
and the simulation with uniform trapping (II). } 
\end{table}

\end{document}